\documentstyle{article}
\def\al{\alpha}
\def\be{\beta}
\def\om{\omega}
\def\dta{\delta}
\def\Dta{\Delta}
\def\sig{\sigma}
\def\ul{\underline}
\def\ket#1{|#1\rangle}
\def\gtoe{{\it g} \leftrightarrow {\it e}}
\def\tvib{\tau_{\rm vib}}
\def\trad{\tau_{\rm rad}}
\def\beq{\begin{equation}}
\def\eeq{\end{equation}}
\def\Bp{B_\perp}

\def\ham{{\cal H}}

\def\bvf{{\bf f}}

\title{Vibrational Decoherence in Ion Trap Quantum Computers
\thanks{To appear in the Proceedings of the Conference on Fundamanetal
Problems in Quantum Theory, University of Maryland Baltimore County,
Baltimore, MD, USA, Aug.~4-7, 1997.}}
\author{Anupam Garg\\
Department of Physics and Astronomy, \\Northwestern University,
\\Evanston, Illinois 60208}
\date{}
\begin{document}

\maketitle

\begin{center}
{\bf Abstract}
\end{center}
The ion trap quantum computer proposed by Cirac and Zoller [Phys. Rev. Lett.
$\ul{74}$, 4091 (1995)] is analyzed for decoherence due to vibrations of the
ions. An adiabatic approximation exploiting the vast difference between the
frequencies of the optical intraionic transition and the vibrational modes is
used to find the decoherence time at any temperature $T$. The scaling of this
decoherence time with the number of ions is discussed, and compared with
that due to spontaneous emission.

\vspace{0.8cm}
\noindent {\bf I. INTRODUCTION}
\vspace{0.1cm}

Since Shor's discovery [1] of an algorithm for factorization of a composite number
of order $2^L$ in $\sim L^3$ steps on an ideal quantum computer (QC), a great
deal of effort has gone into developing quantum computational theory, and related
ideas in information transfer and cryptography. This work has shed new light on
both quantum mechanics and computational complexity theory. As Landauer [2] has so
pungently said, however, writing down a Hamiltonian is not the same as specifying
an apparatus, and it is also necessary, at some time, to look into building real
machines. Landauer's criticism has not gone completely unanswered, and concrete
proposals for implementing QC's have been put forward. Perhaps the most
promising of these is by Cirac and Zoller (CZ) [3] -- but see also Ref. [4]. It
is based on a linear array of trapped ions driven by a precisely timed
sequence of laser pulses. This proposal is being taken seriously enough that at
least one group in the world is trying to build a prototype [5].

The CZ QC consists of $N$ identical ions, trapped and cooled in a linear rf Paul
trap. The ions form a linear array with nonuniform spacings determined by their
mutual Coulomb repulsion and the effective trapping potential (see Fig. 1).
Two internal states of each ion, $\ket g$ and $\ket e$, serve as the quantum bit,
and laser pulses ($\pi$, $\pi/2$, etc.) drive $\gtoe$ transitions and thus implement
one-bit gates. These transitions can be driven either by lasers tuned to the direct  
$\gtoe$ transition frequency $\om_0$, or by Raman pulses in a $\Lambda$ system,
where the lower levels form the quantum bit, and are chosen from the ground
multiplet of the ion to minimize spontaneous emission. The innovative idea due to
CZ is to use the center of mass longitudinal vibrational mode of the array as
a bus that enables the execution of two-bit gates. This can be done by a combination
of one-ion pulses and pulses detuned by $\om_z$, the above mentioned
vibrational mode frequency, to
any pair of ions. Any superposition of the $2^N$ states of the QC (i.e., the
internal states of the ionic system) can then be converted to any other
superposition by a suitable sequence of one- and two-bit gates. 

In this article we shall analyze the decoherence in the Cirac-Zoller (CZ)
QC. A brief description of this work has appeared elsewhere [6].
It hardly needs to be said that decoherence is a serious limitation to the
functioning of any QC. We will be interesed in the {\it intrinsic} decoherence.
Technical difficulties, such as trapping a large enough number of ions, proper
shaping, phase locking, and timing of the laser pulses, optical resolution of
individual ions etc., also have the same practical effect as decoherence, and may
well turn out to be insurmountable by themselves, but that is a separate matter.
We will mainly discuss the decoherence from the vibrations of the ions. Since
the discussion in the earlier paper [6] was terse and technical, we will
focus in the present article on the physical explanation of why ionic vibration
is decohering, and give a qualitative estimate of the decoherence time.
Other
discussions specific to the CZ scheme are by Plenio and Knight [7], and by Hughes
et al. [8]. General discussions of why and how decoherence is detrimental to QC's 
have been given by Landauer [2] and by Unruh [9], among others.

Intrinsic decoherence in the CZ QC arises from two sources: spontaneous emission
and ionic vibration. It is easy to estimate the decoherence rate from the first.
If the spontaneous $\ket e \to \ket g$ decay time for one ion is $\tau_s$, and we
assume that $N/2$ ions are in the excited state on average, we obtain an upper
bound of $\trad \approx 2\tau_s/N$ for the window of time in which any
computation must be completed, since a {\it single} spontaneous decay irretrievably
disrupts the wavefunction of the QC as a whole. Our estimate ignores effects like
superradiance, or changes in the decay of one ion due to the proximity of the other
ions, but this is justified if the inter ion spacing is larger than $2\pi c/\om_0$,
the wavelength of the $\gtoe$ spectral line.

A similar estimate of $\tvib$, the computional time limit imposed by
vibrational decoherence, is not so easy to obtain. Whatever it is, one simple
point should be noted now. Since the two mechanisms of decoherence operate
independently and in parallel, we should add their {\it rates} to obtain the
total decoherence limit on the useful working time of the QC:
\beq
t_{\rm dec} = \left( \trad^{-1} + \tvib^{-1} \right) ^{-1}.
\eeq
If it should happen that one of the rates, $\trad^{-1}$ and $\tvib^{-1}$,
is much larger than the other, this
would help us relax the design constraints, as we could then ignore the slow decay
process to a first approximation. The faster decay may of course still overwhelm us.

The plan of the paper is as follows. The calculation of $\tvib$  is done in Sec.~II.
The results are discussed in Sec.~III. Certain mathematical details are relegated
to two appendices.

\vspace{0.8cm}
\noindent {\bf II. VIBRATIONAL DECOHERENCE}
\vspace{0.1cm}

The physical origin of vibrational decoherence is as follows. Suppose ion $j$
(see Fig. 1) is not in its equilibrium position. It creates an excess electric
field (or electric field gradient) on a neighboring ion, $i$. This excess field
alters the evolution in the $\ket e$, $\ket g$ space of ion $i$ from the desired
time evolution, and the accumulation of this effect causes a decay in the
probability that the QC will be in the intended state. 

To qualitatively estimate
the decay time, let us denote the longitudinal position of the $j$th ion by $z_j$,
and the deviation from this position by ${\bf u}_j$. Let us further suppose that the
$\gtoe$ transition is of electric dipole (E1) or quadrupole (E2) type, and denote
the relevant transition matrix elemenent by $d_a$, where the index $a=1$, or 2,
for the E1 and E2 cases respectively. Ignoring vector and tensor indices, and
denoting the ionic charge by $q$, the change in the $eg$ matrix element of the
Hamiltonian for ion $i$ is given by
\beq
\dta V_i = d_a\sum_{j\ne i} {q\over |z_i - z_j|^{a+2}} u_j.
\eeq

The key point now is not only that this perturbation is small, i.e.,
$|\dta V_i| \ll \hbar\om_0$, but that it is also slow. It varies over times
set by the periods of the normal modes of the ion array, which are
much longer than the optical transition time $\om_0^{-1}$. In other words,
\beq
\left| {d\over dt}\ln |\dta V_i| \right| \ll \om_0.
\eeq
The slowness enables us to treat the perturbation $\dta V_i$ adiabatically. Let 
us map each two-state ion onto a spin-1/2, with $\ket e$ and $\ket g$ being the up
and down spin states. The $i$th spin then sees magnetic fields $B_z = \hbar\om_0$
and $\Bp = \dta V_i$ (see Fig. 2). Since $B_z \gg |\Bp|$, the instantaneous
precession frequency for this spin is given by
\begin{eqnarray}
\om'_{0i} & = & (\om_0^2 + \dta V_i^2/\hbar^2)^{1/2} \nonumber \\
          & \approx & \om_0 + {\dta V_i^2 \over 2\hbar^2 \om_0}.
\end{eqnarray}
On the other hand, the precession axis for the spin can be taken to be
$\hat{\bf z}$ at all times to very good approximation. Thus the time evolution
of the spin up and spin down states $\ket\pm$ is given by
$\exp\left(\pm i\int_0^t dt'\,\om'_{0i}(t')/2 \right) \ket{\pm}$.
A more formal derivation of this result is given in Appendix A.

The phase of the $i$th
ion thus wanders off course by $\pi$ in a time
$\tau_i \approx \pi/(\om'_{0i} - \om_0)$. Using Eqs.~(2) and (4), we can write
\beq
\tau_i^{-1} \approx {q^2d_a^2 \over 2\pi\hbar^2\om_0}
            \left\langle \biggl(
            \sum_{j\ne i} {u_j \over (z_i -z_j)^{a+2}} \biggr) ^2
                                                      \right\rangle .
\eeq
The angular brackets above denote some kind of average. The motion of different
ions is correlated via the normal modes, which we can describe by a density matrix
at some temperature $T$. (This is what one means by the statement that the ions
have been cooled to a temperature $T$.) Additional correlations due to the systematic
manipulation of the center of mass mode used to execute the two-bit gates should not
be included since this is part of the designed time evolution and does not represent
a decay. It is therefore completely consistent to have left out this part of the
time evolution in arriving at Eq.~(5). To obtain the best-case answer, we will assume
that the ion temperature $T \ll \hbar\om_z/k_B$, and approximate this by $T=0$.
To obtain an order of magnitude, we ignore the details of the normal modes, and simply
take
\beq
\langle u_j u_k \rangle = {\hbar \over m \om_t} \dta_{jk},
\eeq
where $\om_t$ is a typical transverse mode frequency, and $m$ is the ionic mass.
This yields
\beq
\tau_i^{-1} \approx {q^2d_a^2 \over 2\pi\hbar m \om_0\om_t}
                    \sum_{j\ne i}{1\over (z_i - z_j)^{2a + 4}}.
\eeq

The rationale for using a transverse mode frequency above is two-fold. First, these
frequencies are generally higher than the longitudinal ones, and second, it is best
to choose the states $\ket g$ and $\ket e$ to be such that the longitudinal modes
cannot excite any transitions due to a $J_z$ selection rule.

The next step is to obtain the decoherence rate $\tvib^{-1}$ for the QC as a whole.
One's first guess would be that this is obtained by just adding the $\tau^{-1}_i$
for all $i$. The detailed calculation of Ref.~[6] shows that this is not quite
correct. The correct procedure is to add the squares and then take the square root. In
other words,
\beq
\tvib^{-2} = \sum_i \tau_i^{-2}.
\eeq
A qualitative justification for this formula is as follows. The overlap between the
actual and intended states of the {\it i\/}th spin is better approximated by
$\cos(t/\tau_i)$ instead of $\exp(-t/\tau_i)$. [See Eq.~(21).]
The probability $P(t)$ that the
QC is in the desired state is thus approximately $\prod_i\cos^2(t/\tau_i) \simeq
\exp(-t^2/\tvib^2)$ with $\tvib$ as given above. 
[If we simply added $\tau_i^{-1}$, we would obtain an additional factor of
$N^{1/2}$ in the scaling behavior of $\tvib^{-1}$, and Eq.~(11) below, e.g., would
contain an $N$ instead of the $N^{1/2}$.]

By combining Eqs. (7) and (8), we obtain a formal answer for the decoherence rate.
The sums over the lattice positions $z_i$, are at worst, numerical problems.
When $N\gg 1$, however, we can evaluate these sums by invoking a continuum
approximation for the array. This approximation and the lattice sums are discussed
in Appendix B. The final result can be written as
\beq
\tvib^{-1} \sim N^{1/2} {q^2 d_a^2 \over 2\pi\hbar m\om_0\om_t s_0^{2a+4}}.
\eeq
where $s_0$ is the minimum spacing between the ions which occurs at the center of
the array. We can make it apparent that
$\tvib^{-1}$ is a rate by noting that [see Eq.~(23)], $q^2 = m\om_z^2d_0^3$, where
$d_0$ is the the trap length scale parameter, and that
\beq
d_a^2 \propto  \hbar /\tau_s k_0^{2a+1},
\eeq
where $k_0 = \om_0/c$. It follows that
\beq
{1\over \tvib} \sim {N^{1/2} \over \tau_s} \biggl({d_0 \over s_0} \biggr)^3
                 {\om_z^2 \over \om_0\om_t} 
                {1 \over (k_0 s_0)^{2a+1}}.
\eeq

Discussion of this result and comparison with the spontaneous decay rate is
given in the next Section.

\vspace{0.8cm}
\noindent {\bf III. DISCUSSION}
\vspace{0.1cm}

To apply Eq.~(11) we must take into account that
the scaling of $\tvib^{-1}$ with $N$ depends critically on how the trap
operating conditions are varied with $N$, and can not be naively taken as $N^{1/2}$.
If, for example,  $s_0$ is held fixed as $N$ is increased, then
$(d_0/s_0)^3 \sim N^2/\ln N $ and $\tvib^{-1} \sim N^{5/2}/\ln N$. In this case,
however, the longitudinal voltage on the trap electrodes, which is proportional
to $\om_z^2$, varies as $(\ln N)/N^2$. This leads to an increase in the total
computational time, since the time for executing a primitive two-bit gate varies
as $\om_z^{-1}$, i.e., as $N$. Also, since the longitudinal
trapping is weaker, the ion array becomes
more susceptible to patch voltages on the electrodes, and non-linear effects in the
trapping potential become more important. If, on the other hand, the
trap voltages, and therefore, $\om_z$ and $\om_t$, are held fixed as $N$ increases,
then $\tvib^{-1} \sim N^{(8a+19)/6}(\ln N)^{-(2a+4)/3}$, i.e.,
as $N^{9/2}(\ln N)^{-2}$ for an E1 transition, and as $N^{35/6}(\ln N)^{-8/3}$ for
an E2 transition. In this case, the minimum spacing $s_0$ varies as
$\sim N^{-2/3}$, and it may become difficult to resolve the ions optically as is
necessary to execute the basic gates. 
Obviously, any intermediate variation is possible by allowing
both $s_0$ and $\om_z$ to change with $N$, and the exact manner in which this is
done is thus a
matter of detailed engineering considerations, which it is premature to discuss.  

To get a numerical estimate of $\tvib$, we will consider the case of
Ba$^+$ ions, which are particularly favorable from the standpoint of minimizing
spontaneous emission decoherence. We choose for $\ket g$ a state in the ground multiplet
$6s\, ^2S_{1/2}$, and for $\ket e$ a state in the first excited multiplet
$5d\,^2 D_{5/2}$. The frequency $\om_0 = (2\pi)1.7\times 10^{14}$ Hz.
Since $\Dta L =2$, the $^2D_{5/2} \to\  ^2S_{1/2}$ decay is an
E2 process, and the spontaneous decay time is $\tau_s = 35\,$s [10]. (There is 
some uncertainty over this number. Hughes et al. [8] take it as 47 s, and it may
even be as high as 70 s. We have taken an average from Ref. [10].) The index $a$
is $2$. Further, we take $\om_z/2\pi = 100\,$kHz, and
$\om_t/2\pi = 20\,$MHz. Then $d_0 = 14\,\mu$m, and for $N=1000$,
$\tvib \simeq 10^4 \tau_s$, which is surprisingly large. (It is even larger in
comparison to $\trad = \tau_s/N$.) On the other hand, $s_0 \simeq 0.5\,\mu$m with the
same parameters, which runs into the difficulty with optical resolution mentioned
above. This suggests that a compromise in which $\om_z$ is reduced may work better
but we have not explored this point further.

That $\trad$ is so much longer than $\tvib$ for the example chosen above
may make one wonder if one should not have anticipated this fact. We do not believe
so. The situation would change completely for larger $N$. Indeed for large enough
$N$, our calculation shows that vibrational decoherence will always dominate over
spontaneous decay decoherence. Secondly, the scaling with $N$ is quite non-trivial
and unexpected.

We conclude that ionic vibrations are not a significant source of decoherence in
the original scheme envisaged by CZ, at least for $N\le 1000$. This should not be
taken to mean that the problems posed by radiative decoherence by themselves are not
serious. Indeed, this could well make the whole scheme unworkable. Similarly, the
challenges of trapping 1000 ions, and of addressing them individually are at the
moment quite daunting. Nevertheless, our conclusions are encouraging in that they
enable us to focus on the radiative decay problem. Several authors [7,8] have
suggested working with a $\Lambda$ system and Raman pulses as a way of dealing with
this. An evaluation of the vibrational decoherence in this setup remains to be done.

\vspace{0.6cm}
\noindent {\bf ACKOWLEDGMENTS}\newline
{}\newline
\indent This work is supported by the National Science Foundation via Grant No.
DMR-9306947.

\vspace{0.8cm}
\noindent{\bf APPENDIX A: SPIN-$1/2$ IN SLOW AND WEAK TRANSVERSE FIELD}
\vspace{0.1cm}

We consider in this Appendix a spin-1/2 system with the following Hamiltonian
\beq
\ham = {1\over 2}\om_0 \sig_z + {\bvf}(t)\cdot{\vec\sig},
\eeq
where ${\bvf}(t) = (f_x,f_y,0)$ has no $z$ component. Further $\bvf$ is small and
slow as explained in Sec. II. (The field $\Bp$ introduced
in Sec. II is just twice $\bvf$.) We are interested in solving for the time
evolution of an arbitrary state for general $\bvf (t)$.

Denoting the eigenstates of $S_z$ with eigenvalues $\pm 1/2$ by $\ket\pm$,
let us write a general state of the spin as
\beq
\ket{\psi(t)} = u_+(t) e^{-i\om_0t/2}\ket+  + u_-(t) e^{i\om_0t/2} \ket{-}.
\eeq
Schr\"odinger's equation then takes the form
\beq
i{\dot u}_{\pm} = e^{\pm i\om_0 t} f_{\mp}(t) u_{\mp}(t),
\eeq
where $f_\pm = f_x \pm i f_y$.

Since $f_\pm$ varies very slowly, we seek the answer to Eq.~(14) in the form
\beq
u_\pm(t) = \al_\pm(t) +\be_\pm(t),
\eeq
where $\al_\pm$ and $\be_\pm$ are fast and slow parts, the latter varying little over a
period $2\pi/\om_0$, and the former averaging to zero over several such periods.

Substituting Eq.~(15) in (14), and separating the fast and slow parts, we obtain
\begin{eqnarray}
i{\dot \al}_{\pm} & = & e^{\pm i\om_0 t} f_{\mp}(t) \be_{\mp}(t), \\
i{\dot \be}_{\pm} & = & e^{\pm i\om_0 t} f_{\mp}(t) \al_{\mp}(t).
\end{eqnarray}
To integrate Eq.~(16), it is a good approximation to
treat the slowly varying functions $f_\pm$ and $\be_\pm$
as constants. In this way, we obtain
\beq
\al_{\pm} = \mp \om_0^{-1} e^{\pm i\om_0 t} f_{\mp}(t) \be_{\mp}(t).
\eeq
We now put this solution in Eq.~(17), and average the resulting equation
over several periods $2\pi/\om_0$. This yields
\beq
i{\dot \be}_\pm = \pm {|\bvf(t)|^2 \over \om_0} \be_\pm (t).
\eeq
Integrating this, we obtain $\be_\pm (t) = \exp(\mp i\Phi(t)) \be_\pm(0)$, where
\beq
\Phi(t) = \int_0^t dt'\, {|\bvf(t')|^2 \over \om_0}.
\eeq
We can take $u_\pm \approx \be_\pm$, since $|\al_\pm| \ll |\be_\pm|$.
The quantity $\Phi(t)$ is precisely the excess angle through which the spin precesses
about the ${\bf{\hat z}}$ axis as discussed in Sec. II, i.e., it is half the
difference between $\int_0^t \om'_{0i}(t')\,dt'$ and $\om_0t$.

Suppose the initial state of the spin is $2^{-1/2}(\ket+ + \ket-)$, i.e.,
$u_\pm (0) = 2^{-1/2}$. Let us denote the state at time $t$ that would be
obtained in the absence of the transverse field $\bvf$ by $\ket{\psi_0(t)}$.
In the context of the QC, the states $\ket{\psi_0(t)}$ and $\ket{\psi(t)}$ are
analogous to the states of the ideal and actual QC, without and with decoherence,
respectively. The extent of decoherence is given by the overlap
\beq
\langle\psi_0(t) | \psi (t) \rangle = \cos(\Phi(t)).
\eeq

A more careful calculation shows that $\Phi(t)$ also has a Berry phase part
$\hat{\bf z} \cdot (\bvf \times {\dot\bvf})$. For the problem of interest to us,
this is much smaller then the dynamical phase and may be neglected. 
Further, the adiabatic approximation breaks
down due to secular effects for $t \sim \om_0^3/|{\dot\bvf}|^2$. It is not hard to
see that this breakdown time is much larger than $\tvib$, and so the approximations
(4) or (20) are completely adequate. 

\vspace{0.8cm}
\noindent {\bf APPENDIX B:  CONTINUUM APPROXIMATION FOR ION ARRAY}
\vspace{0.1cm}

We wish in this Appendix to quantitatively understand the structure of the linear
array of trapped ions when $N \gg 1$. We do this via a continuum approximation
based on the expectation that the local spacing $s(z_i)$ between ions in the
vicinity of ion $i$ will vary slowly with $i$.  Our goal is to find the function
$s(z)$, where we regard $z$ as a continuous variable. We will also find how
the total length of the array, $2L$, varies with $N$.

A simple-minded argument for $s(z)$
is as follows.  Consider the Coulomb forces on an
ion at position $z$ from its nearest neighbours to the left and right, which
we take to be at positions $z-s_-$ and $z+s_+$ respectively. The net force is then
$q^2(s_-^{-2} - s_+^{-2}) \approx 2q^2s^{-2} (ds/dz)$, where we have approximated
$s_+ - s_-$ by $s(z) (ds/dz)$. The force from successively distant pairs of
neighbours is smaller than this expression by factors of 4, 9, 16, etc., since
the distances are
approximately doubled, tripled and so on. Thus the net Coulomb force is
$(\pi^2q^2/3s^2)(ds/dz)$, since $\sum_n n^{-2} = \pi^2/6$. Equating this to the
opposing spring force $m\om_z^2 z$ from the trapping potential, we obtain
\beq
{\pi^2 \over 3 s^2(z)}{ds \over dz} = {z\over d_0^3},
\eeq
where
\beq
d_0 = (q^2/m\om_z^2)^{1/3}
\eeq
is a natural length scale for the trap. (It is easy to show that the ion
spacing is of the order of $d_0$ for 2 or 3 ions in the trap.)
Denoting the total length of the array by $2L$ and placing the center at $z=0$,
integration of Eq.~(22) gives
\beq
{1\over s(L)} - {1\over s(z)} = - {3\over 2\pi^2d_0^3} (L^2 - z^2).
\eeq
By balancing the forces on the ion at the end of the array in the same way as
was done above, we obtain $s(L) \approx \pi (d_0^3/6L)^{1/2}$. We can thus ignore
$s^{-1}(L)$ compared to $3L^2/2\pi^2d_0^3$ in Eq.~(24), which yields
\beq
s(z) = s_0 (1-z^2/L^2)^{-1},
\eeq
where $s_0 \equiv s(0) = 2\pi^2d_0^3/L^2$ is the minimum ion spacing (attained
at $z=0$).

To obtain an expression for $L(N)$, let us denote the ion number at position $z$
by $n(z)$. Then, integration of the approximate relation $dn/dz = 1/s(z)$ using
Eq.~(25) gives $L=d_0(\pi^2 N/2)^{1/3}$. This implies $s_0 \sim N^{-2/3}$. The
mean spacing can be found to vary as $N^{-2/3} \ln N$.

It is clear that the above argument does not treat the ends of the chain properly,
and also underestimates the Coulomb forces due to the more distant neighbors. A
more sophisticated approach is due to Dubin [11], who treats the ion array as a
fluid of total charge $Nq$. In a harmonic trapping potential, the solution to this
problem is known: the fluid forms a blob of uniform charge density in
the shape of an ellipsoid of revolution of
total volume $4\pi N d_0^3$ provided $\om_t \gg \om_z$. If
the semi major axis of this ellipsoid is $L$, the semi minor axis is therefore
$(3Nd_0^3/L)^{1/2}$. The inverse spacing $1/s(z)$ is clearly the number of 
charges per unit length along the major axis of the ellipsoid, and this in turn is
given by its cross-sectional area. In this way we obtain
\beq
{1\over s(z)} = {3\over 4}{N\over L}\biggl( 1-{z^2 \over L^2} \biggr).
\eeq
This is identical to our approximate form (25), but $s_0$ is different.

We still need to find
the length $L$. The fluid approximation breaks down over here, as the answers it
gives depend on the ratio $\om_z/\om_t$, which is clearly wrong as long as the
linear structure is stable. Dubin uses a local density functional theory to
estimate the correction to the Coulomb energy due to the discreteness of the array,
and minimizes the sum of this correction, the fluid drop self energy, and the
trapping potential energy, with respect to $L$. The result for $L$ is then
independent of $\om_z/\om_t$ and is given by
\beq
L^3 = 3N \ln(0.8N) d_0^3.
\eeq
(The 0.8 is actually $6e^{\gamma - 13/5}$, with $\gamma$ being Euler's constant.)
This result differs from our previous one by logarithmic factors. We also obtain
\beq
s_0 = 4L/3N = 1.92 N^{-2/3} [\ln(0.8 N)]^{1/3} d_0.
\eeq
This result should also be compared to that of
Hughes et al.~[8], who find on the basis
of a numerical fit that $s_0 = 2 N^{-0.56} d_0$. Since for moderate values of $N$, the
logarithmic factor in Eq.~(28) will have the effect of increasing the apparent
exponent of $N$, these two results are quite comparable. We do not know if Hughes et
al. did numerics for large enough $N$ to discern the presence or absence of
logarithmic factors, so it is hard to say which is better.

The above results can be used to perform
the sums over lattice positions that appear in Eqs.~(7) and (8).
There are two types of sums. The first,
\beq
S_n(i) \equiv \sum_{j\ne i}{1\over |z_i - z_j|^n},
\eeq
can be very simply evaluated as
\beq
S_n(i) \approx {2 s^n(z_i)} \sum_{j=1}^\infty {1\over j^n}
       = {2\zeta(n) \over s^n(z_i)}.
\eeq
This approximation is actually fairly good for all $i$ except very close to
the ends, since the exponent
$n$ is at least 3 or 4 in all cases that we encounter. 

The second type of sum is $T_n = \sum_i s^{-n}(z_i)$. This can be approximated
by an integral. Writing $\Delta i \approx dz/s(z)$, we obtain
\beq
T_n = \sum_i {1\over s^n(z_i)} \approx \int\limits_{-L}^L {dz\over s^{n+1}(z)}.
\eeq
With $s(z)$ given by Eq.~(25), the integral is elementary, and we obtain
\beq
T_n \approx {L\over s_0^{n+1}} \biggl( {4\pi \over 4n + 7} \biggr)^{1/2},
\eeq
where the last form comes from an asymptotic formula for $\beta(n+2,1/2)$.

\vspace{0.8cm}
\noindent {\bf REFERENCES}
\newcounter{refs}
\begin{list}
   {[\arabic{refs}]}{\usecounter{refs}}

\item{P. Shor, in {\it Proceedings of the 35th Annual Symposium on the Foundations
of Computer Science} (IEEE Computer Society, Los Alamitos, CA, 1994.)}
\item{R. Landauer, Philos. Trans. R. Soc. London A $\ul{353}$, 367 (1995).}
\item{J. I. Cirac and P. Zoller, Phys. Rev. Lett. $\ul{74}$, 4091 (1995).}
\item{T. Pellizzari, S. A. Gardiner,  J. I. Cirac, and P. Zoller, Phys. Rev. Lett.
$\ul{75}$, 3788 (1995).}
\item{R. Hughes, these proceedings.}
\item{Anupam Garg, Phys. Rev. Lett. $\ul{77}$, 964 (1996), atom-ph/9603009.}
\item{M. B. Plenio and P. L. Knight, Phys. Rev. A $\ul{53}$, 2986 (1996).}
\item{R. J. Hughes {\it et al.}, Phys. Rev. Lett. $\ul{77}$, 3240 (1996).}
\item{W. Unruh, Phys. Rev. A $\ul{51}$, 992 (1995).} 
\item{D. A. Church, Phys. Rep. $\ul{228}$, 253 (1993). See Table 7.b.}
\item{D. H. E.  Dubin, Phys. Rev. Lett. $\ul{71}$, 2753 (1993).}
\end{list} 

\pagebreak
\noindent LIST OF FIGURES
\begin{description}
\item[Fig. 1.] Schematic of the array of trapped ions in the
Cirac-Zoller quantum computer.

\item[Fig. 2.] Equivalent magnetic fields acting on the
internal states of the {\it i\/}th ion in the spin representation.
\end{description}
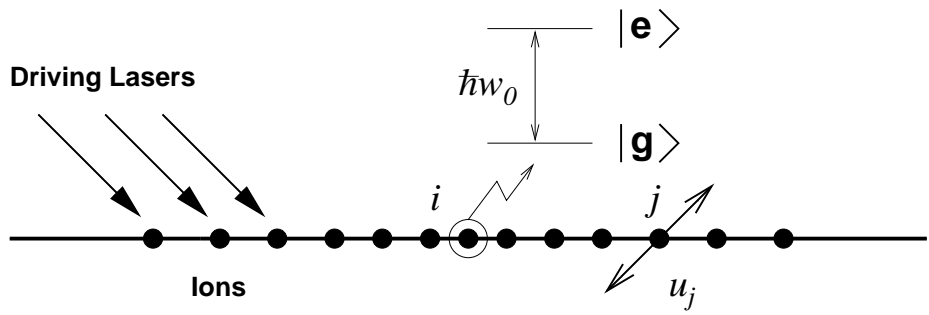
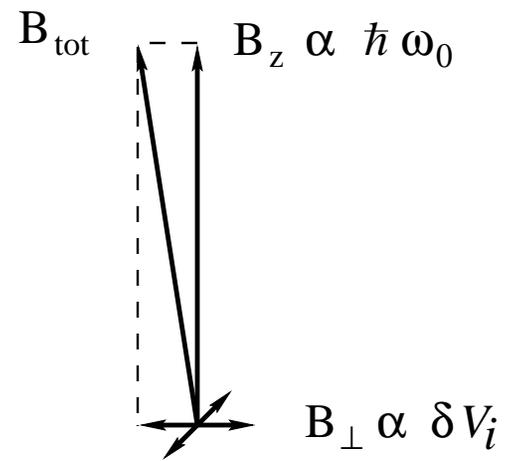
\end{document}